\documentclass[%
 reprint,
superscriptaddress,
 amsmath,amssymb,
 aps,
prl,
]{revtex4-1}

\usepackage{graphicx}
\usepackage{dcolumn}
\usepackage{bm}

\begin{document}

\preprint{APS/123-QED}

\title{
Is $\gamma$-ray emission from novae affected by interference effects in the $^{18}$F(p,$\alpha$)$^{15}$O reaction?}

\author{A. M. Laird\footnote{corresponding author: alison.laird@york.ac.uk}}
\affiliation{Department of Physics, University of York, York, YO10 5DD, UK} 
\author{A. Parikh}
\affiliation{Departament de F\'{i}sica i Enginyeria Nuclear, EUETIB, Universitat
Polit\`{e}cnica de Catalunya, c/ Comte d’Urgell 187, E-08036 Barcelona,
Spain}
\affiliation{Institut d’Estudis Espacials de Catalunya, c/Gran Capita 2-4, Ed.
Nexus-201, E-08034 Barcelona, Spain}
\author{A. St.J. Murphy}
\affiliation{SUPA, School of Physics and Astronomy, The University of Edinburgh, Edinburgh, EH9 3JZ, UK}
\author{K. Wimmer}
\affiliation{National Superconducting Cyclotron Laboratory, Michigan State
University, East Lansing, Michigan 48824, USA}
\affiliation{Department of Physics, Central Michigan University, Mt. Pleasant, Michigan, 48859, USA}
\author{A. A. Chen}
\affiliation{Department of Physics and Astronomy, McMaster University, Hamilton,
Ontario, Canada L8S 4M1}
\author{C.~M.~Deibel}
\affiliation{Physics Division, Argonne National Laboratory, Argonne, Illinois 60439, USA}
\affiliation{Joint Institute for Nuclear Astrophysics, Michigan State University,
East Lansing, Michigan 48824, USA}
\author{T. Faestermann}\affiliation{Physik Department E12, Technische Universit\"{a}t M\"{u}nchen, DE-85748
Garching, Germany}
\affiliation{Maier-Leibnitz-Laboratorium der M\"{u}nchner Universit\"{a}ten (MLL),
DE-85748 Garching, Germany}

\author{S. P. Fox}
\affiliation{Department of Physics, University of York, York, YO10 5DD, UK}

\author{B.~R.~Fulton}
\affiliation{Department of Physics, University of York, York, YO10 5DD, UK}
\author{R. Hertenberger}
 \affiliation{Fakult\"{a}t f\"{u}r Physik, Ludwig-Maximilians-Universit\"{a}t M\"{u}nchen, DE-85748, Garching, Germany}
\affiliation{Maier-Leibnitz-Laboratorium der M\"{u}nchner Universit\"{a}ten (MLL),
DE-85748 Garching, Germany}
\author{D.~Irvine}
\affiliation{Department of Physics and Astronomy, McMaster University, Hamilton,
Ontario, Canada L8S 4M1}
\author{J.~Jos\'{e}}\affiliation{Departament de F\'{i}sica i Enginyeria Nuclear, EUETIB, Universitat
Polit\`{e}cnica de Catalunya, c/ Comte d’Urgell 187, E-08036 Barcelona,
Spain}
\affiliation{Institut d’Estudis Espacials de Catalunya, c/Gran Capita 2-4, Ed.
Nexus-201, E-08034 Barcelona, Spain}
\author{R. Longland}
\affiliation{Departament de F\'{i}sica i Enginyeria Nuclear, EUETIB, Universitat
Polit\`{e}cnica de Catalunya, c/ Comte d’Urgell 187, E-08036 Barcelona,
Spain}
\affiliation{Institut d’Estudis Espacials de Catalunya, c/Gran Capita 2-4, Ed.
Nexus-201, E-08034 Barcelona, Spain}
\author{D. Mountford}
\affiliation{SUPA, School of Physics and Astronomy, The University of Edinburgh, Edinburgh, EH9 3JZ, UK}
\author{B. Sambrook}
\affiliation{Department of Physics and Astronomy, McMaster University, Hamilton,
Ontario, Canada L8S 4M1}
\author{D. Seiler} \affiliation{Physik Department E12, Technische Universit\"{a}t M\"{u}nchen, DE-85748
Garching, Germany}
\affiliation{Maier-Leibnitz-Laboratorium der M\"{u}nchner Universit\"{a}ten (MLL),
DE-85748 Garching, Germany}
\author{H.-F. Wirth}
\affiliation{Fakult\"{a}t f\"{u}r Physik, Ludwig-Maximilians-Universit\"{a}t M\"{u}nchen, DE-85748, Garching, Germany}
\affiliation{Maier-Leibnitz-Laboratorium der M\"{u}nchner Universit\"{a}ten (MLL),
DE-85748 Garching, Germany}

\date{\today}

\begin{abstract}
The $^{18}$F(p,$\alpha$)$^{15}$O reaction rate is crucial for constraining model
predictions of the $\gamma$-ray observable radioisotope $^{18}$F produced in novae. The determination of this rate is challenging due to particular features of the level scheme of the compound nucleus, $^{19}$Ne, which result in interference effects potentially playing a significant role. The dominant uncertainty in this rate arises from interference between J$^{\pi}$=3/2$^+$ states near the proton threshold (S$_p$ = 6.411 MeV) and a broad J$^{\pi}$=3/2$^+$ state at 665 keV above threshold. This unknown interference term results in up to a factor of 40 uncertainty in the astrophysical S-factor at nova temperatures. Here we report a new measurement of states in this energy region using the $^{19}$F($^3$He,t)$^{19}$Ne reaction. In stark contrast with previous assumptions we find at least 3 resonances between the proton
threshold and E$_{cm}$=50 keV, all with different angular distributions. None 
of these are consistent with J$^{\pi}$= 3/2$^+$ angular distributions. We find that the main uncertainty now arises from the unknown proton-width of the 48 keV resonance, not from possible interference effects.
Hydrodynamic nova model calculations performed indicate that this unknown width affects $^{18}$F production by at least a factor of two in the model considered.

\begin{description}
\item[PACS numbers]
26.50.+x, 26.30.Ca, 25.55.Kr
\end{description}
\end{abstract}

\maketitle

Novae  
occur in binary systems where hydrogen-rich material is accreted from a companion star onto a white dwarf, leading to thermonuclear runaway and subsequent ejection of material. Their ejecta is thought to be the main source of $^{13}$C, $^{15}$N and $^{17}$O in the Galaxy \cite{Starr,Jos07}. The relevant unstable nuclei are accessible to experiments, and consequently, novae are the only explosive environment where the nuclear physics input is almost entirely based on experimental data \cite{Jos06}. 

However, there are a number of outstanding challenges
in our understanding of nova explosions \cite{Jos08}, one of which is to reproduce 
the amount of ejected material inferred from infrared and radio observations, 
which is systematically underestimated by models. 
An independent way to constrain the ejected masses would be the detection of 
$\gamma$-rays, produced at the explosion stage. When the envelope becomes
optically thin, novae are expected to emit $\gamma$-rays,   
dominated by a prominent 511 keV line. 
Predicted detectability distances of this prompt $\gamma$-ray emission (about 2 - 3 kpc \cite{Jos07})
strongly depend on the overall amount of $^{18}$F (T$_{1/2}$($\beta^+$)=110 mins) left over after the explosion. This is critically influenced by the $^{18}$F(p,$\alpha$)$^{15}$O reaction. 
Sensitivity studies of the impact of reaction rates on nova
nucleosynthesis suggest that rates should be known to a precision of, at least, 30\%
\cite{Jos06}. However, this rate is currently poorly understood and considerable experimental and theoretical effort has been focused 
on determining this rate (\cite{Beer,Nesa} and references therein). 

Until recently, this rate was thought to be dominated by (i) the 3/2$^-$  resonance at E$_{cm}$ = 330 keV, 
 and (ii) the interference of the 3/2$^+$ states, at 8 and 38 keV E$_{cm}$, with the known, broad 3/2$^+$ resonance at 665 keV. The cross section in the astrophysically important energy region can vary by up to a factor of 40 for different
assumptions of the interference terms \cite{Beer}.
This interference contribution cannot be calculated but must be measured in the relevant energy range. A predicted broad 1/2$^+$ sub-threshold state \cite{DD,Dal,DM} could also contribute significantly in the region of interest if present.

The 330 keV resonance corresponds to a 3/2$^-$ state \cite{Garrett,Utku,Visser} at E$_x$ = 6.741 MeV in $^{19}$Ne.  The contribution of this resonance to the 
$^{18}$F(p,$\alpha$)$^{15}$O cross section has been measured directly by Bardayan {\em et al.} \cite{Bar02} and Beer {\em et al.} \cite{Beer}. 

The situation regarding the  J$^{\pi}$  = 3/2$^+$ states is less clear. Two states at 8 and 38 keV E$_{cm}$ were
first observed by Utku {\em et al.} \cite{Utku} via the $^{19}$F($^3$He,t)$^{19}$Ne reaction. They were both tentatively assigned to be J$^{\pi}$  = 3/2$^+$ but no explanation for this was given. A compilation by Nesaraja {\em et al.} \cite{Nesa} states that these assignments are based on similarities in excitation energy and the small energy shift expected compared to analogue states in the mirror nucleus, $^{19}$F. 

Recent results using the $^{18}$F(d,n)$^{19}$Ne reaction \cite{Adek}, however, suggest that these assignments may be incorrect. The 8 keV resonance was observed and the measured angular distribution 
indicated a J$^{\pi}$ assignment of 1/2$^-$, 3/2$^-$ or 5/2$^-$ \cite{Adek,DB}. However, the 38 keV resonance was not observed. Crucially, if the 8 keV resonance is not considered to be 3/2$^+$ then the argument regarding mirror states, made in \cite{Nesa}, no longer applies, and the J$^{\pi}$ of the 38 keV resonance is experimentally unconstrained.
A sub-threshold state observed at -122 keV (E$_x$ = 6.289 MeV) was considered to be either a 1/2$^+$ or 3/2$^+$ state. Although this state is far below the proton threshold and not broad enough to contribute directly, a 3/2$^+$ assignment would lead to interference with the broad 3/2$^+$ resonance at 665 keV.

It follows that the cross section in the region between the proton threshold and the 330 keV resonance, and thus the $^{18}$F(p,$\alpha$)$^{15}$O reaction rate at nova temperatures, is now poorly constrained. Improved spectroscopic information is needed, particularly on the location of the crucial J$^{\pi}$  = 3/2$^+$
states, to allow the possible effects of interference on the reaction rate to be determined. Moreoever, the experimental approach adopted must not only populate these states, but also provide sufficient resolution to 
separate states assumed to be only 30 keV apart. Of the studies performed to date, only that of Utku {\em et al.} \cite{Utku} provided clean population of the states of interest with resolution close to that required. As the original tentative 3/2$^+$ assignments also arose from that work, a re-measurement allows these assignments to be re-evaluated.

In this Letter, we report a study of the level structure of $^{19}$Ne 
through the $^{19}$F($^3$He,t)$^{19}$Ne reaction. The reaction was studied at the Maier-Leibnitz-Laboratorium (MLL) in Garching, Germany, using the same method and equipment previously reported in \cite{Parikh}.  A 25 MeV beam of $^3$He$^{2+}$ ions was 
delivered to the target position of a quadrupole-dipole-dipole-dipole (Q3D) magnetic spectrograph \cite{Loff}.  Targets included a 50 $\mu$g/cm$^2$ CaF$_2$ deposited upon a 7 $\mu$g/cm$^2$ foil of enriched  $^{12}$C, and a 25 $\mu$g/cm$^2$ aluminum foil.  Measurements were made at spectrograph laboratory angles between 10$^{\circ}$ and 50$^{\circ}$.
Tritons from ($^3$He,t) reactions on contaminants, including $^{12}$C and $^{16}$O, were excluded from the focal-plane detector \cite{Wirth} by virtue of their Q-values.

\begin{figure}[htb]
\centering
\includegraphics[scale=0.5, angle=0]{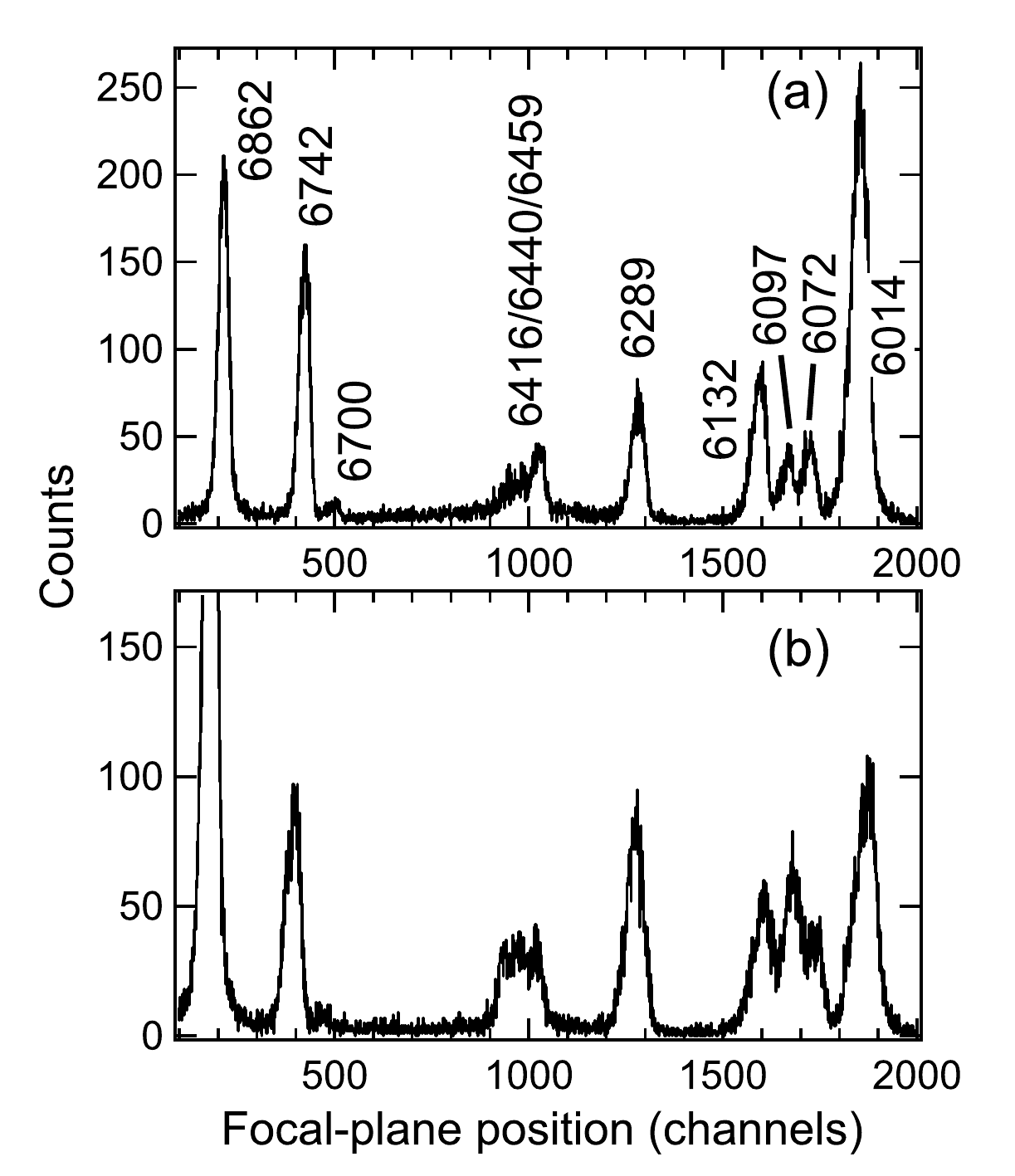}
\caption{\small Raw focal-plane triton spectra at $\theta_{lab}$ = 10$^{\circ}$ (a) and 20$^{\circ}$ (b). Excitation energies are labeled in keV.}
\label{tritons}
\end{figure}

\begin{figure}[htb]
\centering
\includegraphics[scale=0.6, angle=0]{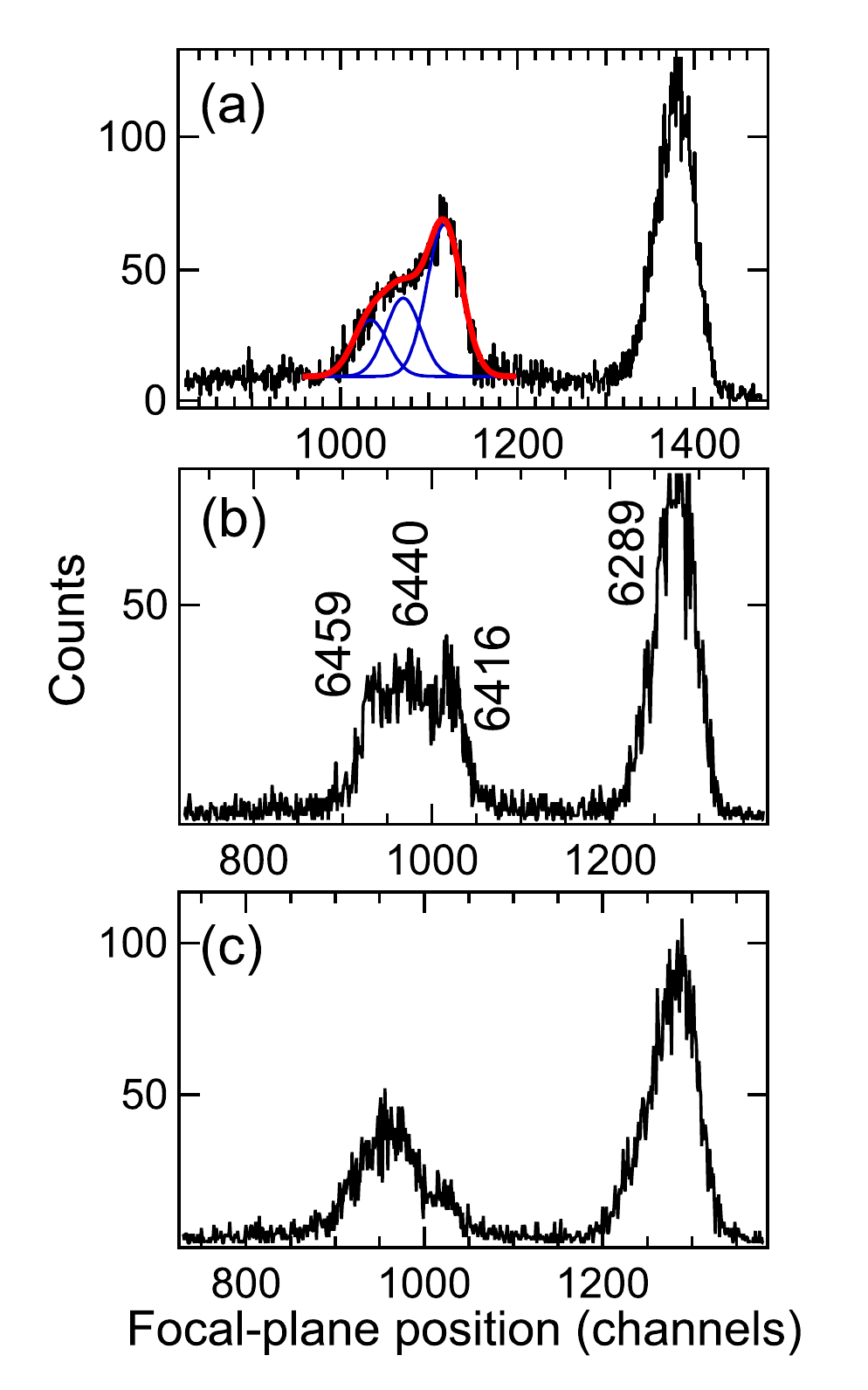}
\caption{\small (Color online) Partial raw focal-plane triton spectrum at $\theta_{lab}$ = 15$^{\circ}$ (a), 20$^{\circ}$  (b) and 30$^{\circ}$ (c). At 15$^{\circ}$, the overall best fit (red online) and three constituent Gaussian peaks (blue online) are shown for the states within E$_x$ = 6.4 - 6.5 MeV.}
\label{zoom}
\end{figure}

Fig. \ref{tritons} shows triton position spectra  
at angles of 10$^{\circ}$ and 20$^{\circ}$.  These spectra were analysed using least-squares fits of multiple Gaussian or exponentially-modified Gaussian functions with a constant background. 
Peak widths were fixed to $\sim$ 14 keV FWHM based on fits of isolated peaks in the spectra. Fig. \ref{zoom} shows partial focal-plane spectra at 15, 20 and 30 deg, highlighting our observation of three states between 6.4 and 6.5 MeV.  

At each angle the focal-plane was calibrated using well-resolved, known states in $^{27}$Si \cite{Endt,Lotay}, with 4.2 $<$ E$_x$($^{27}$Si) $<$ 5.5 MeV, populated via the $^{27}$Al($^3$He,t) reaction.  Second-degree polynomial fits of triton radius-of-curvature to focal-plane position channel were obtained at each angle, and these fits were used to determine excitation energies for states 
 in $^{19}$Ne (e.g., Fig. \ref{tritons}).  Those energies corresponding to clearly resolved, strongly populated  states in each spectrum were later used as part of an internal calibration to determine the energies of the three states between E$_x$($^{19}$Ne) = 6.41 - 6.46 MeV.       

Excitation energies from this work are listed in Tab. \ref{resparamtab}.  These energies are all weighted averages of energies determined from at least four different measurement angles.
In addition, we note a systematic uncertainty of $\pm$ 2 keV due to the uncertainty in the thicknesses of the Al and CaF$_2$ targets and the uncertainty in the relative Q-value of the $^{19}$F($^3$He,t)$^{19}$Ne and $^{27}$Al($^3$He,t)$^{27}$Si reactions \cite{Audi}.  

Between E$_x$ (E$_{cm}$) = 6.41 - 6.46 MeV (0 - 50 keV), the position spectra at each angle require that  three narrow states contribute to the observed feature, rather than the previously assumed two levels at 8 and 38 keV. This feature is best reproduced using energies of 5, 29 and 48 keV, with our assumed line shape. 

\begin{figure}[htb]
\centering
\includegraphics[scale=0.4, angle=0]{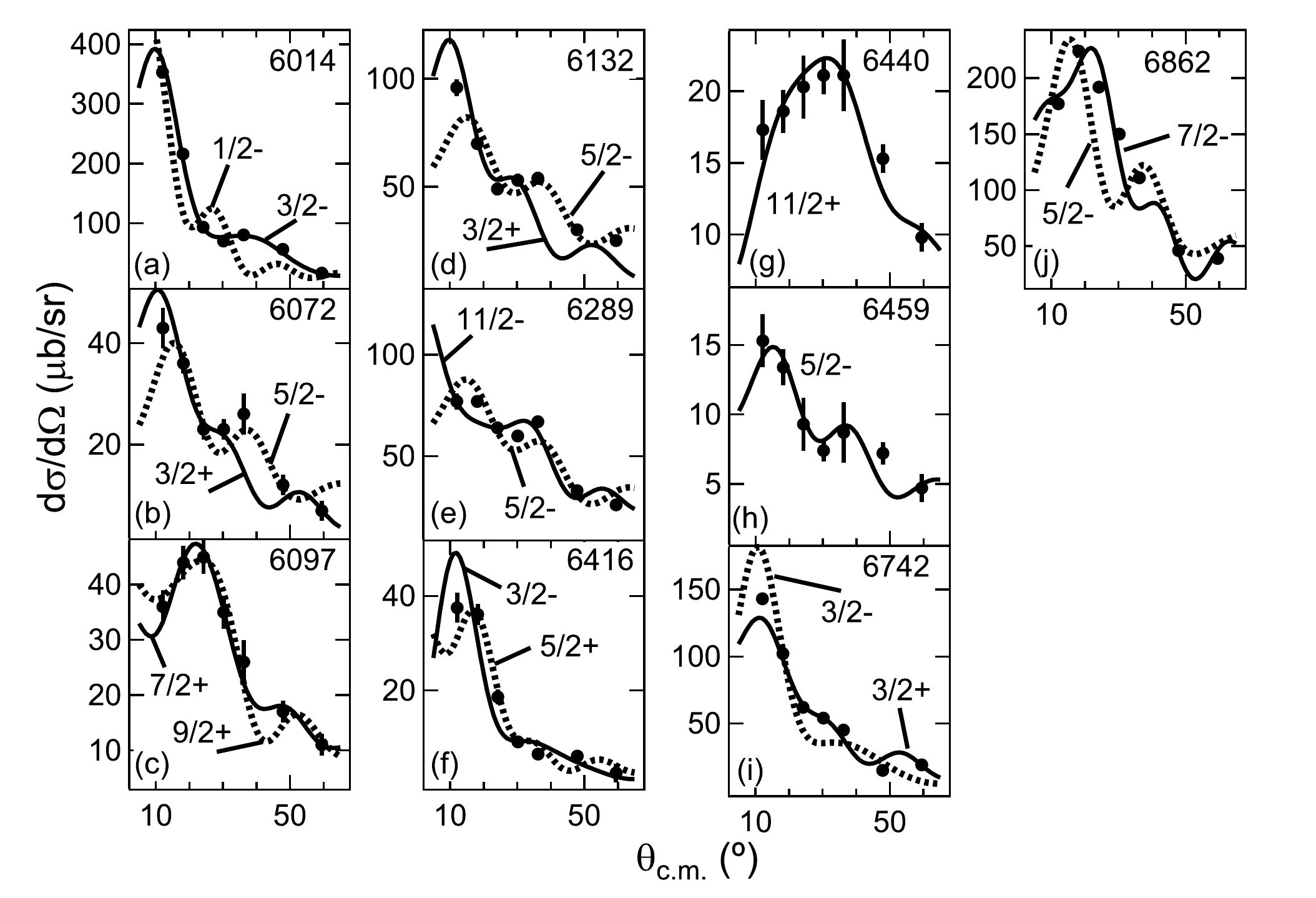}
\caption{\small Triton angular distributions measured with the
$^{19}$F($^3$He,t)$^{19}$Ne reaction.  Curves calculated with 
FRESCO have been fit to the data.
Each panel (a - j) is labeled with the excitation energy (in keV) of
the associated state in $^{19}$Ne and the J$^{\pi}$ values of the curves that
best fit the data.}
\label{fresco}
\end{figure}

By comparing the shapes of the measured angular distributions given in Fig. \ref{fresco}, it is clear that the three narrow states between E$_x$ = 6.41 - 6.46 MeV all have different J$^{\pi}$ values.  The states at 6.014, 6.072, 6.132, 6.416, 6.459 and 6.742 MeV exhibit similar, forward-peaked, angular distributions, suggestive of low spin states. The states at 6.097, 6.289 and 6.862 MeV have similar features in their angular distributions which indicate that these are not low spin. The 6.440 MeV state does not have a forward peaked distribution suggesting that it is higher spin. 
These statements were determined purely from visual inspection of the experimental angular distributions. 

The most important result from these data is that there is now clear evidence that the previously assumed 8 and 38 keV resonances cannot both be 3/2$^+$, and consquently the J$^{\pi}$ of the latter is unknown. 

Distorted Wave Born Approximation calculations were performed, using the finite-range coupled-channels reaction code FRESCO [7].  The ($^3$He,t) charge exchange reaction has been treated as a two-step ($^3$He,d)(d,t) reaction. This method allows the extraction of the angular momentum transfer of the reaction, since the shapes of the angular distributions are very similar to those from the one step ($^3$He,t) reaction \cite{Pind,Gaarde,Parikh}. 
The optical model parameters have been taken from \cite{Ver82} for the $^3$He+$^{19}$F entrance channel, \cite{Ver94} for the intermediate $^2$H+$^{20}$Ne channel and \cite{Perey} for the exit channel $^3$H+$^{19}$Ne. The FRESCO angular distributions, also shown in Figure \ref{fresco}, provide some level of quantitative constraint on the  J$^{\pi}$ assignments. These are summarised in Tab. \ref{resparamtab} and particular cases discussed below. 

\begin{figure}[htb]
\centering
\includegraphics[scale=0.4, angle=0]{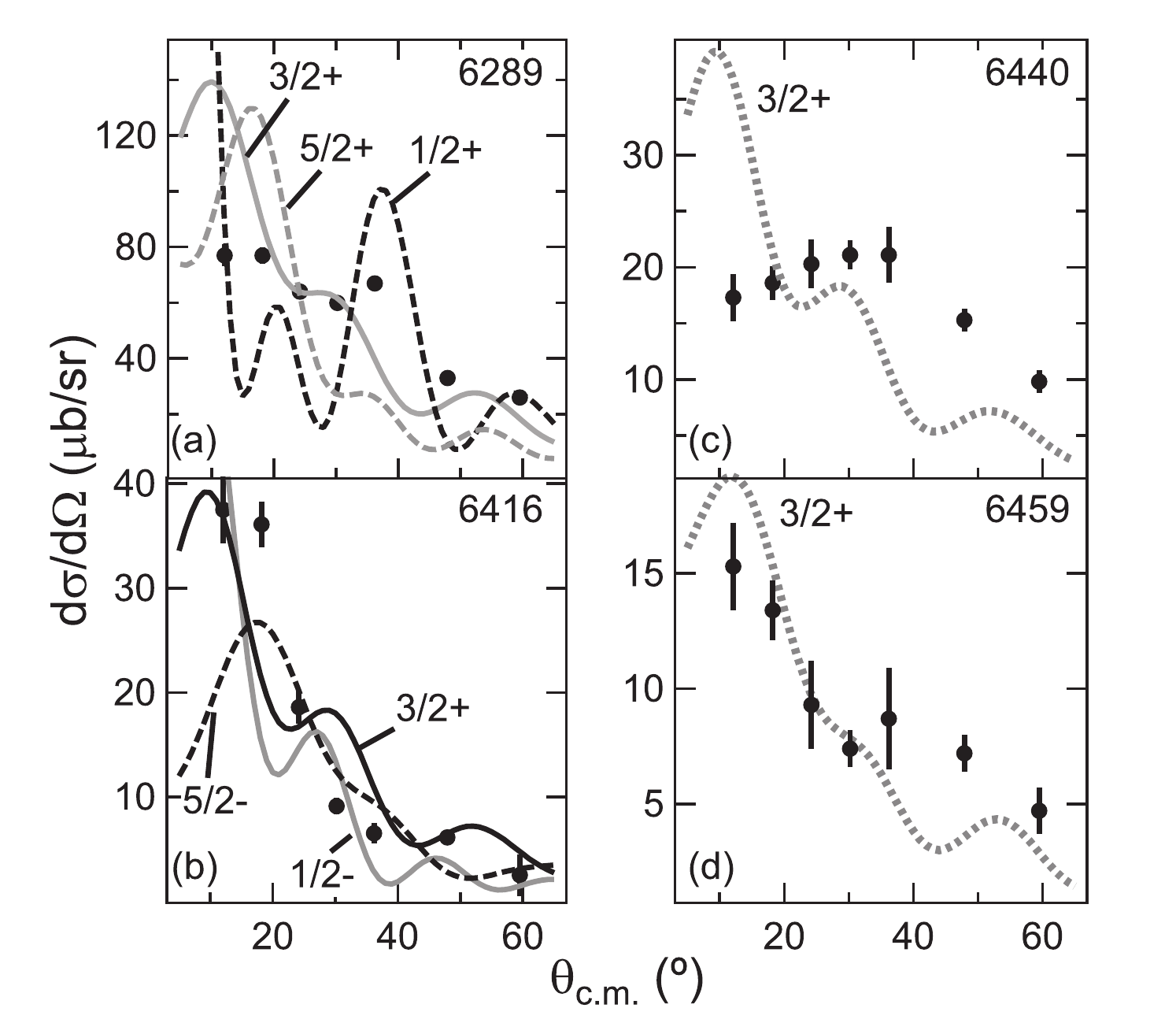}
\caption{\small Triton angular distributions measured with the
$^{19}$F($^3$He,t)$^{19}$Ne reaction at 25 MeV.  Curves calculated as in Fig \ref{fresco}. 
Fits shown are for previously accepted J$^{\pi}$ assignments.}
\label{fresco_z}
\end{figure}

Crucially, of the three states just above the proton-threshold, none are found to be consistent with a 3/2$^+$ assignment. Fig. \ref{fresco_z} shows the angular distributions for these three states again, this time compared to FRESCO calculations for 3/2$^+$ and other previously assumed J$^{\pi}$ assignments. Also shown is the sub-threshold state at 6.289 MeV \cite{Adek}.
The 6.416 MeV (5 keV) state is consistent with either 3/2$^-$ or 5/2$^+$. The 6.440 MeV (29 keV) state is clearly not reproduced by a 3/2$^+$ assignment and best fit with an 11/2$^+$. This assignment is supported by the compilation of \cite{Nesa} which lists an expected 11/2$^+$ state in this region. For the 6.459 MeV (48 keV) state, the 3/2$^+$ calculation cannot reproduce the low and high angle data simultaneously and so is excluded. A 5/2$^-$ assignment best reproduces the data. Finally, the 6.289 MeV state is not well reproduced by any calculation, but those with high spin (>3/2) are preferred.  This state may be an unresolved doublet. While the reaction mechanism for the ($^3$He,t) is complex, the reasonable reproduction of known J$^{\pi}$ 
assignments gives confidence in the assignments from the calculations. Additional theoretical study of the ($^3$He,t) reaction at these energies would be very valuable
Results from this work are compared in Tab. \ref{resparamtab} with recent studies of $^{19}$Ne states. 

The indication that none of the states around 6.4 MeV seem to be consistent with 3/2$^+$ raises the question of the possible location of these 3/2$^+$ states observed in the mirror and
whether they contribute significantly to the $^{18}$F(p,$\alpha$)$^{15}$O rate under nova
conditions. From the present work and previous data, there is no indication of any 3/2$^+$ states between 6.4 and 6.86 MeV. In the sub-threshold region, the states at 6.072 and 6.132 MeV are possible candidates should they prove to be 3/2$^+$ rather than 5/2$^-$. However, as these states are well below threshold it is not clear whether they would play any role in the $^{18}$F(p,$\alpha$)$^{15}$O rate at nova temperatures.

\begin{table*}[htb]
\begin{center}
\begin{tabular}{|c|c|c|c|c||c|c|c|}
\hline
\multicolumn{5}{|c||}{Present work} & \multicolumn{3}{|c|}{Previous work$^{\dagger}$}\\
E$_x$ (MeV) & E$_{CM}$ (keV)	&  	J$^{\pi}$ & $\Gamma_p$ [keV]$^{\ddagger}$ & $\Gamma_{\alpha}$ [keV]	$^{\ddagger}$	&	Ex (MeV) & E$_{CM}$ (keV)	&  	J$^{\pi}$ 	\\
\hline

6.014(2) &  -397 & 3/2$^-$ &- & -&  6.016 &-395  & (1/2,3/2)$^-$ \\

6.072(2) & $^{\star}$ -339 &  (3/2$^{+}$,5/2$^-$) & 0.143& 6$\times$10$^{-4}$ & 6.078 & -333 & - \\
6.097(3) & -314 & (7/2,9/2)$^{+}$ & -& -&  6.107 & -304 & - \\

6.132(3) &  $^{\star}$ -282  & (3/2$^{+}$,5/2$^-$) & 0.143  &7$\times$10$^{-4}$ & 6.138 & -276 & - \\

6.289(3) & -122	& -	& - & - & 6.290 & -121 & (1/2, 3/2,5/2)$^{+}$ \\

6.416(3) & $^{\star}$ 5  &  (3/2$^{-}$,5/2$^{+}$)&	4.7$\times$10$^{-50}$, 1.2$\times$10$^{-51}$	&  0.5, 0.126  &	6.419(6)	&   8 & (1/2,3/2)$^-$   	\\
6.440(3)& 29 	  &	(11/2$^+$)	& -& -& -	& - &	 -  \\
6.459(3) & $^{\star}$ 48 &  5/2$^-$  & 8.4$\times$10$^{-14}$  & 5.5  & 6.449(7) & 38 & (3/2$^{+}$)\\

6.700(3) & 289				&	-	& & & 6.698(6) & 287	
	&	(5/2$^{+}$)		\\

6.742(2) & $^{\star}$ 331  		&	3/2$^{-}$ &  2.22$\times$10$^{-3}$ $^{\natural}$ & 	5.2 $^{\natural}$	&  6.741(6)
& 	330 & 	3/2$^{-}$	\\
6.862(2) & 451 				&	7/2$^{-}$ &		1.1$\times$10$^{-5}$ $^{\natural}$& 1.2$^{\natural}$		&		6.861(6)
	& 450	&	7/2$^{-}$	\\		
\hline
\end{tabular}
\end{center}
\caption{Resonance parameters from the present work compared to previous values.
One should consider an additional systematic uncertainty of +/- 2 keV (see text).
$^{\star}$ used in R-matrix calculations.
$^{\ddag}$ deduced but not measured in the present work, unless otherwise indicated (see text).
$^{\natural}$ parameter taken from \cite{Adek}.
$^{\dag}$ taken from \cite{Utku,Adek,Nesa}. } 
\label{resparamtab}
\end{table*}

To evaluate the impact of the new J$^{\pi}$ information, multi-channel R-matrix calculations were performed using the DREAM code \cite{DREAM}, including 7 states. For the 6.416 MeV state, it was assumed that this corresponded to the state previously identified at 6.419 MeV \cite{Adek} with J$^{\pi}$ = 3/2$^-$. The proton- and alpha-partial widths were recalculated under the assumption of unchanged reduced widths. The widths for the state at 6.459 MeV (here 5/2$^-$) were recalculated for the change of energy and angular momentum from the parameters of \cite{Nesa} for the 6.449 MeV state. The reduced proton width was calculated to be 0.014, which was considered to be in line with similar states in this region. For the sub-threshold states, the widths were deduced assuming the same reduced widths as found by \cite{Adek} for their sub-threshold $\ell$=0 state, and a  J$^{\pi}$ = 3/2$^+$. The parameters of the states included are given in Tab. \ref{resparamtab}. The contribution of the 6.440 MeV was negligible due to the high spin, and so was not considered. 

By comparing the individual reaction rates
of the 5 and 451 keV resonances with that of the 331 keV resonance, it is clear that neither plays any significant role in nova explosions. 
The 331 keV resonance dominates between 0.25 and 0.4 GK, with the broad 665 keV resonance dominating between 0.1 and 0.25 GK, and above 0.4 GK. Using the assumed reduced proton-width of 0.014 for the 48 keV resonance, it is the dominant contribution below $\sim$ 0.12 GK only.  However, this proton width is, in practice, unconstrained. Taking a realistic upper limit of its reduced proton-width to be 0.1, the 48 keV resonance would dominate up to around 0.25 GK, i.e. over a significant part of the nova temperature range.

The total reaction rate was then calculated assuming no 3/2$^+$ sub-threshold states, hereafter our nominal rate. Initial calculations demonstrated that only 3/2$^+$ sub-threshold states produced a significant contribution above threshold. Therefore, upper and lower (constructive and destructive) interference rates, assuming the 6.072 and 6.132 MeV states to be 3/2$^+$, were also calculated to establish the impact of the uncertainty in the parameters of the sub-threshold states. Finally to explore the uncertainty arising from the unknown proton width of the 48 keV resonance, two additional rates were calculated, based on the nominal rate but assuming firstly zero contribution from the 48 keV, and secondly a reduced proton width of 0.1 (rather than 0.014). 

Using these reaction rates, 5 hydrodynamic nova simulations were performed.
We have adopted a typical case consisting of a 1.15 M$_\odot$ ONe white dwarf,
accreting solar material at 2$\times10^{-10}$ M$_\odot$.yr$^{-1}$ and assuming 50\% mixing
between accreted material and the outermost ONe substrate.
The final $^{18}$F
abundances were compared one hour after peak temperature (0.23 GK). The upper and lower interference rates show a 50\% abundance decrease and increase, respectively, compared to the nominal rate.  
The uncertainty associated with the 48 keV resonance, however, results in a factor of $\sim$ 2 uncertainty in the final $^{18}$F yield, which in turn,
affects the predicted maximum detectability distance for the associated $\gamma$-ray lines by about a factor 1.4 for the models considered.

In conclusion, a study of $^{19}$Ne states has been performed using the $^{19}$F($^3$He,t)$^{19}$Ne reaction.  Angular distributions were measured for ten states between 6.0 and 6.9 MeV E$_x$. The feature in previous data at $\sim$ 6.3 MeV E$_{x}$ assumed to be due to a 3/2$^+$ pair of states has been shown to consist of three states, all with different J$^{\pi}$.
DWBA calculations, using a two-step assumption, suggest that none are consistent with 3/2$^+$. 
Reaction rates have been calculated for the possible J$^{\pi}$ permutations and corresponding uncertainties in $^{18}$F abundance determined. The largest rate uncertainty now arises from the unknown proton width of the 48 keV resonance. Therefore experimental efforts should be made to confirm the location and J$^{\pi}$ of this resonance and determine its proton width. Determination of the J$^{\pi}$ of the two sub-threshold states would also aid in the reduction of the uncertainty in the $^{18}$F(p,$\alpha$)$^{15}$O reaction rate.

\begin{acknowledgments}
The authors would like to thank the MLL staff for their support during the setup and running of the experiment. UK personnel were supported by the Science Technology Funding Council (STFC). This work was supported by the DFG Cluster of Excellence  "“Origin and Structure of the Universe”"
(www.universe-cluster.de). AP, JJ and RL were partially supported by the
Spanish MICINN under Grants No. AYA2010-15685 and No.
EUI2009-04167, by the E. U. FEDER funds, and by the ESF
EUROCORES Program EuroGENESIS. AAC was supported,
in part, by a grant from NSERC Canada. CMD
acknowledges support from the US Department of Energy, Office of Nuclear Physics, under Contract No. DE-AC02-
06CH11357. CMD was also partially supported by JINA
Grant No. PHY0822648.
\end{acknowledgments}

\end{document}